\documentclass{mn2e}
\usepackage{psfig}
\usepackage{mnras_cite}
\newcommand{\centralhi}{\left( \frac{n_{o}}{0.1 \, {\rm cm^{-3}}} \right)}

\newcommand{\Egas}{\left( \frac{kT_{e}}{8 \, {\rm keV}} \right)}
\begin{document}

\title[Excess Ionization and Soft X-ray Emission from Cooling Flow Clusters]{Excess Ionization and Soft X-ray Emission from Cooling Flow Clusters} 
\author[Oh] 
{S. Peng Oh\\
Dept. of Physics, University of California Santa Barbara, Santa Barbara, CA 93106, USA.}

\maketitle

\begin{abstract}
X-ray spectroscopy of cooling-flow clusters reveal an unexpected
deficit of soft X-ray emission lines from gas at $\sim 1/3$ of the
ambient plasma temperature, across a wide range of X-ray luminosities
and virial temperatures. We propose excess ionization from either a
population of suprathermal electrons or photoionization by X-ray continuum emission from
hot gas or a central AGN as a means of decoupling the thermal state of
the gas from its emission line spectrum. The former effect is thought to operate in the solar corona. Because they generally become
important at some fixed fraction of the cluster gas temperature, such
mechanisms could in principle explain both the universality and temperature
dependence of the emission line suppression, properties which none of the present-day models based on gas heating can explain. Ultimately these models cannot explain the observations; however, they have attractive and robust features which could be
useful in elucidating a final solution to the soft X-ray deficit. 
\end{abstract}

\section{Introduction}

Cooling flow clusters have long been a mystery. While the X-ray
luminosity is consistent with gas cooling rates of up to $100-1000
{\rm M_{\odot} \, yr^{-1}}$, whether such a cooling flow is indeed
  really operating and the fate of the cooled gas has been a matter of
  long-standing controversy (see \scite{Fabian1994} for a review). X-ray spectroscopy acquired by the Reflection Grating
Spectrometer on XMM-Newton has added yet another puzzle to the
picture: there is a strong deficit of emission lines associated with the
low temperature gas which should be present in a cooling flow. In particular, the observed differential emission measure from clusters is consistent with \cite{petersonetal2002}:
\begin{equation}
\frac{d EM}{dT}= \frac{5}{2}\frac{\dot{M}}{\mu m_{p}}
\frac{k_{B}}{\Lambda(T)} (\alpha +1) \left( \frac{T}{T_{o}}
\right)^{\alpha}
\label{eqn:emission_measure}
\end{equation}
where $\dot{M}$ is the inferred mass dropout rate, $T_{o}$ is the
background temperature, and $\alpha \approx 1-2$. This conflicts with
the standard isobaric cooling model, which predicts $\alpha=0$,
independent of the details of gas clumping or geometry (\scite{Fabian1994} and references therein). The
observed emission measure of low temperature gas is therefore lower than
expected. In particular, while the spectra exhibits strong emission
from the ambient temperature $T_{o}$ down to about $T_{o}/2$--which
suggests that cooling is indeed taking place--there is
signifiantly less emission from lower temperature gas than is
expected. What is particularly striking is that this seems to be a
universal phenomenon, across a wide range of cluster temperatures and
mass deposition rates: emission from gas at $\sim T_{o}/3$ is always
strongly suppressed. This hints at some common underlying physical
mechanism which scales with the plasma temperature.

A whole host of explanations have been put forth (for reviews,
see \scite{fabianetal2001}, \scite{kaastraetal} and references
therein). Many center on heating the gas, to prevent it from cooling
below some minimum temperature (e.g. through AGN outflows, electron thermal
conduction, mergers, or magnetic reconnection). This suffers from two
generic problems: firstly, it is notoriously difficult to heat gas in a
stable manner; for instance, conduction tends to either make the gas
isothermal (contrary to observations), or allow cooling to continue
unimpeded, unless the conduction suppression factor is excessively
fine-tuned \cite{bregmandavid,meiskin88} (though see \scite{kim_narayan} for a countervailing claim). Secondly, such solutions are much more
effective at altering the normalization of the emission measure (by
reducing $\dot{M}$), rather than its shape. Indeed, for consistency with observations, heating
must either completely suppress local thermal stability or heat the
gas all the way back up to the ambient temperature. If heating delays
but does not halt cooling at some lower temperature, this will increase the
emission measure of low temperature gas over predictions of the
standard isobaric cooling model, rather than reduce it. Another
possibility is to cool gas rapidly by some process which does not
involve X-ray emission (e.g. turbulent mixing, dust cooling), though
such proposals to date suffer from similar difficulties in that it is difficult to naturally identify a characteristic temperature scale at which such effects become important (as is required by observations).
%the cooling affects both hot and cool gas equally. 
Other proposals (more similar in spirit to the route
taken in this paper) seek to suppress the emission lines by
resonant scattering or absorption \cite{petersonetal2001}, or by an inhomogeneous metallicity
distribution which significantly alters the predicted emission line
spectrum \cite{morrisfabian}. However, none have proven wholly successful, and the problem remains unsolved.   

In this paper we pursue another avenue: breaking the expected correspondence between the thermal state
of cooling gas and its emission spectrum. If the ionization stage of an atom does not
correspond to its value in collisional ionization equilibrium, then
the emission line spectrum ceases to be a good thermometer of the
plasma temperature. From equation (\ref{eqn:emission_measure}), the
luminosity in an emission line at frequency $\nu$ from an ionization
stage $i$ is given in the standard isobaric cooling flow model (where
$\alpha=0$) by:
\begin{equation}
L_{\nu}= \frac{5}{2}\frac{k_{B}}{\mu m_{H}}{\dot{M}}
\int_{T_{min}}^{T_{max}}
\frac{\epsilon_{\nu}(T,Z)}{\Lambda(T,Z)}{dT}. 
\label{eqn:line_luminosity}
\end{equation}
Note the inverse weighting by the cooling function: emission at lower
temperatures is given progressively less weight as the gas cools
rapidly through that phase. Since $dL_{\nu} \propto \left[ x_{i}(T)/\Lambda(T) \right] dT$ (where $x_{i}$ is the ionization fraction of stage $i$),
if we can delay the appearance of an ionization stage $i$ until low
temperatures when cooling is rapid, the emission measure of its
associated lines will be severely reduced. This could happen, for
instance, if the gas falls out of collisional equilibrium; however,
for conditions in cooling flow clusters $t_{\rm rec} \ll t_{\rm
  cool}$, so this is unlikely. On the other hand, the gas can also be
overionized if there is some additional source of ionization. We consider two
sources: collisional ionization from a non-Maxwellian tail of
suprathermal electrons, and photo-ionization by free-free
emission from hot intracluster gas. Both of these effects are
associated with the large reservoir of hot gas in the cluster, and the
associated ionization scales naturally with the temperature of
the ambient gas. Thus, the onset of excess ionization takes place at some
fraction of the cluster virial temperature. This would naturally explain the universality of the
low temperature cut-off and the reason why it always appears at some
fixed fraction ($\sim 1/3$) of the temperature of the hot
gas, features which none of the presently popular models can explain. For
instance, it is difficult to understand why heating mechanisms like
AGN heating should target low temperature gas at the expense of high
temperature gas, particularly when cool gas occupies a very small
fraction of the cluster by volume.     
 
Despite the above promising characteristics, we find that 
excess ionization fails to solve the soft X-ray cooling flow
problem, for relatively general and model-independent reasons. In
clusters, suprathermal electrons are much more efficient at heating
the gas via Coulomb collisions rather than ionizing the metals (unlike
in the solar corona, where their ionization effects have been invoked
to explain anomolous ionization patterns, e.g. \scite{scudder94}). Photoionization can
certainly produce excess ionization at low temperatures with little
associated photo-heating, since the the ionizing photons only target
the metals, which are a rare tracer species. However, the same excess ionization reduces the efficiency of gas cooling at low temperatures
which ultimately {\it increases} the emission measure of low
ionization stage emission lines. This arises because the missing
emission lines are produced at temperatures where line cooling, rather
than bremstrahhlung, is the dominant cooling mechanism; one therefore
cannot alter the ionization properties of the gas without altering its
cooling properties. Despite the failure of these two attempts, the
generality and temperature scaling of such alterations to standard
isobaric cooling emission spectrum models is very
attractive, and we feel, under-explored. It is possible that some
variant of these ideas may ultimately prove successful. 
 
\section{Collisional Ionization: Suprathermal Electrons}
\label{section:collisional}

Let us consider a scenario where the electron velocity distribution is
not strictly Maxwellian, but has a non-Maxwellian tail of high
velocity electrons. In this case, the usual correspondence between plasma
temperature and metal ionization stage may be broken: while the gas
can cool down to low temperatures, the metals are maintained at a relatively high ionization stage by the non-Maxwellian tail of suprathermal electrons. Such
ionization effects due to a non-Maxwellian tail is thought to take
place in the solar corona (Roussell-Dupre 1980, Scudder \& Olbert
1979, Cranmer 1998), where the electron mean free path is comparable
to the temperature scale height. Furthermore, if the typical energy of electrons in the suprathermal tail scales with the cluster temperature, this would explain why the cooling cutoff scales with the cluster temperature. Below, we build simple toy models in which a small suprathermal tail attenuates the line flux from lower metal ionization stages by the required amounts. Nonetheless, here we show that this promising scenario does not hold for typical conditions in clusters: any non-Maxwellian
tail of electrons which ionizes metals above their equilibrium
ionization stage at a given temperature will cause an unacceptable
amount of Coulomb heating.   

\subsection{Excess Ionization with a Non-Maxwellian Tail}

The presence of a significant population of
suprathermal electrons in cluster plasma is usually ignored. This is justified on the
basis that the mean free path of electrons is generally much smaller
than the temperature scale height. A minimal estimate of the
temperature scale height $L_{\rm T}$ is the Field length $\lambda_{F}$, which can be obtained in
a local stability analysis by assuming balance between conduction and
cooling \cite{Field1965}:
\begin{equation}
\lambda_{F}= \left[ \frac{f \kappa T^{7/2}}{1.2 n^{2} \Lambda(T)}
\right]^{1/2} = 150 f^{1/2} T_{7}^{3/2} n_{-3}^{-1} ({\rm ln}
\Lambda_{c})/40)^{-1/2} {\rm kpc}
\label{field_length}
\end{equation}  
where ${\rm ln}\Lambda_{c}= 29.7 + {\rm ln}(n^{-1/2} T_{6})$ is the
Coulomb logarithm, $\kappa_{S}$ is the Spitzer coefficient for thermal conduction, and $f$ is some unknown parameter which accounts for the reduction
of the electron mean free path (and thus the thermal conductivity) due
to tangled magnetic fields or plasma instabilities. By contrast, the mean free path of
electrons due to Coulomb collisions is:
\begin{equation}
\lambda_{e} = 20 f \Egas^{2} n_{-3}^{-1} \left(\frac{{\rm ln}\Lambda_{c}}{40}
\right)^{-1} {\rm kpc}
\label{mfp}
\end{equation}
Since $\lambda_{e} \ll \lambda_{F}$, it would appear that the
assumption of classical conductivity is valid, and further that suprathermal
electrons cannot play an important role in cooling flows. However, a
non-Maxwellian tail of suprathermal electrons could plausibly
arise in two cases: (i) plasma instabilities such as whistlers \cite{pistnner}
scatter electrons below some energy $E_{\rm crit}$, but are relatively
ineffective at scattering electrons about this energy. Thermal
conduction will be significantly reduced if most of the
heat-bearing electrons at the peak of the Maxwellian distribution have significaantly attenuated mean-free paths. The
Field length $\lambda_{F} \propto 1/f$ and thus the temperature scale height $L_{\rm T}$ will therefore be severely reduced (indeed, sharp temperature drops indicating
conduction suppression have been observed in clusters \cite{ettorifabian}). Nonetheless,
suprathermal electrons with $\lambda_{e} > L_{\rm T}$ can penetrate the cooling gas and ionize
the metals. Indeed, Chandra has observed a sharp temperature drop in gas from 10 to 5 keV on scales of order $\sim 10-15$kpc \cite{mar2000}, comparable or less than the electron mean free path; such conduction suppression processes may be responsible for the survival of such sharp fronts. (ii) Jets or shocks in the intracluster medium introduce a
secondary population of suprathermal/relativistic electrons. Indeed,
the observation of synchrotron radiation from clusters indicates that
such a population of electrons must exist (e.g., \scite{sarazin}). 

\begin{figure}
\psfig{file=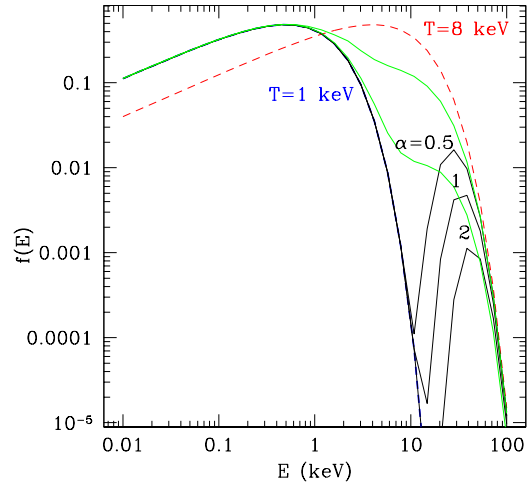,width=80mm}
\caption{The computed electron distribution functions in the cooling clump at 1 keV for 
different $\alpha=H/\lambda$; also shown are Maxwellian distributions for gas at 1 keV an
d 8 keV (the temperature of the ambient ICM). At very high energies, the clump 
is transparent to the hot electrons and the distribution function joins smoothly over to that of the ambient ICM. The non-Maxwellian tail is progressively stronger for smaller clumps (i.e. smaller $\alpha$), which have lower optical depth to Coulomb scattering.} 
\label{fig:DF}
\end{figure}

To demonstrate the effect of such a suprathermal tail on ionization balance, we build a simple toy model. We consider an isobarically cooling filament of gas embedded in a
the ambient ICM at the cluster virial temperature of 8 keV, with initial
density $n \sim 10^{-3} \, {\rm cm^{-3}}$. We
assume a plane-parallel geometry. Our model has one free parameter,
the size of the filament $H$, which we parameterize as
$\alpha=H/\lambda$, where $\lambda \sim 20$kpc is the mean free path
of an 8 keV electron in the ambient ICM. We solve the Fokker-Planck
equation for the electron distribution function, taking into account
the effect of Coulomb collisions. Our approach is similar to that of \scite{owocki_canfield} in the solar corona. We do not attempt to solve 
the full Fokker-Planck equation (e.g. \scite{scoub83}), a computationally
very demanding task. Instead, we use a linearized BGK \cite{krook} approach in which the collision operator is given
by a phenomenological relaxation term. The collisional
thermalization rate is simply given by the rate at which high-velocity test
electrons are deflected by $90^{\circ}$. While in a strict sense the BGK method does not
conserve energy or momentum, in practice it agrees well with
more elaborate calculations \cite{lje88}, and both
the great ease of caculation and the physical insight provided
recommends it for our use. We ignore the effects of a
polarizing electric field (e.g., see Bandiera \& Chen 1994). With these
approximations, we solve:
\begin{equation}
\mu v \frac{\partial f}{\partial z} = \nu(v,z) \left[ f^{*}(v,z) -
f(\mu,z,v) \right]
\label{eqn:DF} 
\end{equation}
where $f$ is the electron velocity distribution function we wish to determine, $f^{*}(v,z)$ is a Maxwellian distribution with temperature $T(z)$, $\nu(r,v) = 16 \pi e^{4} n(r) {\rm ln}\Lambda/(m_{e}^{2} v^{3})$ is the
collisional thermalization rate, and $\mu=v_{z}/v$. This equation is strikingly reminiscent of the equation of
radiative transfer $\frac{dI_{\nu}}{d\tau} = -I_{\nu} + S_{\nu}$, which has
the formal solution $I_{\nu}(\tau_{\nu}) = I_{\nu}(0) e^{- \tau_{\nu}}
+ \int_{0}^{\tau_{\nu}} e^{-(\tau_{\nu}-\tau_{\nu}^{\prime})} S_{\nu}
(\tau_{\nu}^{\prime}) d\tau_{\nu}^{\prime}$. Similar, the solution to equation (\ref{eqn:DF}) consists of a weighted sum of Maxwellians at
different temperature, with the weighting given by the attenuation
${\rm exp}(-\tau)$ due to Coulomb shielding. We show the computed distribution functions for $\alpha=0.5,1,2$ and $\mu=1$ in Figure \ref{fig:DF}. At very high energies, the clump 
is transparent to the hot electrons and the distribution function joins smoothly over to that of the ambient ICM.

\begin{figure}
\psfig{file=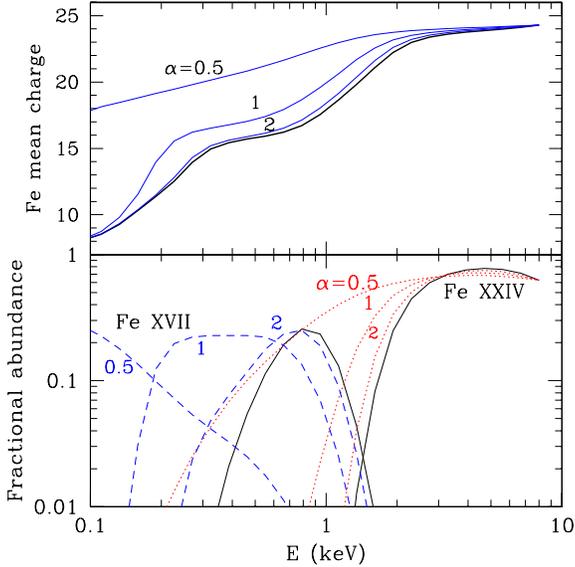,width=80mm}
\caption{{\it Bottom panel}: The fractional abundance of Fe XXIV (whose emission lines are
observed, {\it dotted lines}) and Fe XVII (which emission lines are {\it not} observed, {\it dashed lines}) as
a function of clump temperature. Shown as solid lines are the abundances for
purely Maxwellian electron DFs. For small clumps $\alpha < 0.5$, the
abundance of Fe XVII is strongly suppressed until very low
temperatures, implying that Fe XVII emission lines will not be seen. {\it Top panel}: The mean charge of Fe as a function of clump temperature, for
clumps of different sizes. Shown in black is the expected evolution
for a purely Maxwellian electron DF. For small clumps, suprathermal
electrons keep Fe ions at high ionization stages even at low
temperatures, reducing the usefulness of observed lines as a
temperature diagonostic.} 
\label{fig:ion_frac}
\end{figure}

Given the new distribution function f(v), we compute the new
ionization and recombination rates for Fe ions in the plasma. For a
collisional process with cross section $\sigma (E)$, the rate
coefficient is given by:
\begin{equation}
\Gamma = \left( \frac{2 kT}{m_{e}} \right)^{1/2} \int_{x_{th}}^{\infty}
x^{1/2} \sigma(x kT) f(x) dx  
\end{equation}
where $x_{th}=E_{th}/kT$, and $E_{th}$ is the threshold energy for a
particular process. We obtain the cross sections for all 26 ionization
stages of Fe for collisional ionization (both direct and excitation
autoionization) and recombination (both radiative and dielectronic)
from parametric fits by Arnaud \& Raymond (1992). While the
recombination rates are only mildly affected by the non-Maxwellian
tail of high energy electrons, ionization rates are very significantly
affected, and can be orders of magnitude larger: comparable to the
ionization rate in the ambient medium, for a cooling clump
sufficiently transparent to suprathermal electrons. If we then assume
ionization equilibrium--a good assumption since the recombination time
is very short compared to the cooling time, e.g. for Fe XVII, $t_{rec} \sim 10^{6} n_{-2}^{-1}$yr
$\ll t_{cool} \sim 10^{9} n_{-2}^{-1}$yr--we can then compute the local ionic
abundance $x_{i}$ for all ionization stages as a function of clump temperature. In Figure \ref{fig:ion_frac}, we display the fractional
abundance as a function of local temperature for selected ionization
stages, as well as the mean ionization stage as a function of temperature. Compared to a Maxwellian DF, the mean ionization stage
is significantly higher, particularly at low temperatures. In particular, the
abundance of lower ionization stages as Fe XVII is severely suppressed
until low temperatures, when the cooling time is very short. 

\begin{figure}
\psfig{file=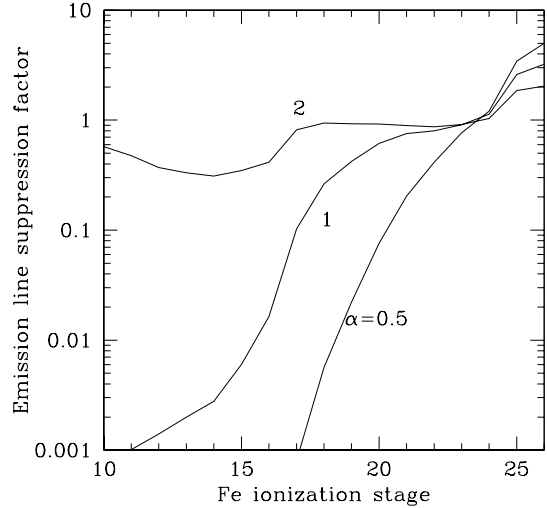,width=80mm}
\caption{Emission line suppression factor $f_{i}^{\rm suppress}$ for each Fe ionization state, as given by equation (\ref{eqn:suppress}), for $\alpha=0.5,1,2$. A purely Maxwellian electron distribution function corresponds to $f_{i}^{\rm suppress}=1$. For distribution functions with a non-Maxwellian tail, line emission from the higher ionization stages is enhanced, while emission from the lower ionization stages is strongly suppressed. For $\alpha=0.5,1$, the suppression far exceeds the minimal amount required to explain observations ($f_{i}^{\rm suppress} \le 0.3$ for Fe XVII and lower in this case).} 
\label{fig:suppress}
\end{figure}

We can quantify the suppression of low ionization stage cooling
lines. Standard isobaric cooling flow prescriptions have an emission
spectrum given by equation (\ref{eqn:line_luminosity}), independent of geometry. The inverse weighting by the cooling function implies that little emission is
associated with gas cooling at low temperatures, when the cooling time
is short. This allows us to compute the suppression factor of line emission
from a particular ionization stage $i$:
\begin{equation}
f^{\rm suppress}_{i} = \frac{\int_{0}^{T_{\rm max}} \frac{x_{i}^{\rm supra}(T)}{\Lambda(T)}}{\int_{0}^{T_{max}} \frac{x_{i}^{\rm maxwell}(T)}{\Lambda(T)}}
\label{eqn:suppress}
\end{equation}
where $x_{i}^{\rm supra}$ and $x_{i}^{\rm maxwell}$ are the fractional
abundances of ions assuming a distribution function with a tail of suprathermal electrons, and a locally Maxwellian DF respectively. We have assumed that the local cooling
rate $\Lambda(T)$ is not significantly affected by suprathermal
electrons (we shall see in \S\ref{section:cloudy} that this assumption is not strictly correct, but here the suppression of the lower ionization stages is so strong that this  correction is unimportant). We therefore adopt a fit to the cooling function of \scite{sutherland_dopita}, for ${\rm Z=0.3 Z_{\odot}}$. We show the results in Figure
(\ref{fig:suppress}), which show emission from lower ionization stages
to be severely suppressed. By comparison, for example, the XMM observations
of Abell 1795 which show a lack of Fe XVII emission lines require that their
emission measure be suppressed by at least a factor of 3 \cite{tamura2001}. Hence, in principle such effects could easily explain the lack of emission lines associated with low temperature gas.  

Note that the suprathermal tail of the electron distribution function in cooling gas is simply given by the high-energy Maxwellian tail of the ambient cluster medium. There is therefore be a
correlation between the virial temperature of the cluster and
ionization stage at which emission becomes strongly suppressed. The cutoff temperature below which line emission is suppressed is always some fraction of the cluster virial temperature. This self-similar feature of the excess ionization model, which closely mirrors what is seen in observations, is a very promising characteristic. It does not emerge naturally in most standard heating models.

%Note that the mean free path of electrons $\lambda \propto T^{2}$,
%while the temperature scale height depends more weakly on the
%temperature of the ambient medium. There should therefore be a
%correlation between the virial temperature of the cluster and
%ionization stage at which emission becomes strongly suppressed. In
%other words, in hotter clusters, the cutoff temperature $T_{\rm min}$ at which gas
%appears to stop cooling should be higher. This is also depicted in
%Figure (\ref{ion_suppress}). 

\subsection{Unacceptable Coulomb Heating}
\label{section:coulomb_heat}

Thus far, we have not self-consistently included the heating due to suprathermal electrons on the temperature profile of the clump. For simplicity let us consider a suprathermal population of electrons
of temperature $T_{h}$ and density $n_{h}$ penetrating a cooling clump
of gas at temperature $T_{c}$ and density $n_{c}$. Note that $T_{h}$
and $n_{h}$ do not necessarily correspond to the temperature and
density of the ambient cluster gas; they are determined by the details
of the electron scattering process. In general $T_{h} > T_{\rm cluster}$
and $n_{h} \ll n_{cluster}$, for high energy electrons far in the
Maxwellian tail. We allow ourselves arbitrary
freedom in the two variables $n_{h},T_{h}$ to see if there exists any
form of the suprathermal tail which would ionize the metals, while still
allowing the gas to cool. 

The rate of energy loss of a highly suprathermal electron $T_{h} \gg
T_{c}$ is independent of $T_{c}$ and given by (Schunk \& Hays 1971):
\begin{equation}
\frac{dT}{dt}=\frac{4 \pi n_{c} e^{4}}{m_{e} v_{e}} {\rm ln} \left(
\frac{m_{e}v_{e}^{3}}{\gamma e^{2} \omega_{p}} \right),
\end{equation}
where $\omega_{p} \equiv 4 \pi n_{c} e^{2} /m_{e}$ is the plasma
frequency. This yields a net Coulomb heating rate:
\begin{equation}
n_{c} \Gamma \approx 5.5 \times 10^{-14} n_{h} n_{c} T_{h}^{-1/2} {\rm erg
\, s^{-1} cm^{-3}}
\end{equation}
The ionization timescale for the ionization stage $i$ is given by
$t_{\rm ion} \equiv n^{i}/\dot{n}^{i}_{\rm ion}=1/n_{h}
\sigma_{i}(v_{h}) v_{h}$ (where $\sigma_{i}(v_{h})$ is the
velocity-dependent ionization cross-section of ionization stage
$i$), while the recombination time is given by $t_{rec}=1/\alpha_{\rm
rec} n_{c}$, where $\alpha_{rec}=\alpha^{\rm rad}_{i+1} + \alpha^{\rm
diel}_{i+1}$ is the sum of the radiative and dielectronic
recombination rates. The ionization cross-section and recombination
rates for Fe are evaluated from parametric fits by Arnaud \& Raymond
(1992). We also define a heating timescale $t_{\rm heat} =
k_{\rm B} n_{c}T_{c}/n_{c} \Gamma$ and a cooling timescale $t_{\rm cool} = k_{\rm B} n_{c}
T_{c}/n_{c}^{2} \Lambda(T_{c})$. 

For definiteness, we consider the effect of suprathermal electrons on
Fe XVII, which is not seen in XMM observations of cooling flow
clusters \cite{petersonetal2001}. We need only consider the temperature range
$T_{c} \approx 0.2-0.8$ keV, where the abundance of Fe XVII becomes
appreciable in a Maxwellian plasma. Since $t_{\rm
rec},t_{\rm cool} \propto 1/n_{c}$, we find:
\begin{equation}
t_{\rm rec} \approx 0.3-1 \times 10^{-3} t_{\rm cool} ; \ \ \ 0.2 {\rm
keV} < T_{c} < 0.8 {\rm keV}
\label{rec_cool}
\end{equation}  
independent of $n_{c}$. Likewise, since $t_{\rm heat},t_{\rm ion}
\propto 1/n_{h}$, we find:
\begin{equation}
t_{\rm ion} \approx 10 t_{\rm heat} \left( \frac{T_{c}}{\rm 1 keV}
\right)^{-1}, 
\label{ion_heat}
\end{equation} 
independent of $n_{h},T_{h}$. The independence from $T_{h}$ arises
because $t_{\rm ion}/t_{\rm heat} \propto \sigma(v_{h}) T_{h}$, and
at high energies the Bethe approximation $\sigma \propto v_{h}^{-2}$
holds. For the abundance of FeXVII to be strongly suppressed, we
require $t_{\rm ion} < t_{\rm rec}$. However, from equations
(\ref{rec_cool}) and (\ref{ion_heat}), this implies 
\begin{equation}
t_{\rm heat} < 10^{-5}-10^{-4} t_{cool}  
\end{equation}
for the relevant temperature range $T_{c} \approx 0.2-0.8$ keV, and
independent of $n_{c},n_{h},T_{h}$. Thus, any population of
suprathermal electrons which is sufficient to ionize Fe XVII at low
temperatures would cause unacceptably large amounts of Coulomb heating
which cannot be radiated away, and
the cooling clump would be rapidly evaporated. Such a large
suprathermal tail would not permit gas to cool significantly below the
ambient cluster temperature to begin with.  This is a robust
conclusion despite our order-of-magnitude approach: the heating rate implied by the required ionization exceeds the gas cooling rate by 4-5 orders of
magnitude. 

Our result can be understood from the fact that the 
cross-section for inelastic collisions of electrons with atoms is much smaller than
the cross-section for elastic Coulomb collisions with other electrons
by at least the value of the fine structure constant, $\sim
1/137$. The effectiveness of frequent small-angle encounters at small
impact parameters decreases the electron-electron equilibration
timescale by another 2 orders of magnitude (Spitzer 1978). Thus, the
plasma is heated much more rapidly by suprathermal electrons than it is
ionized.\footnote{For similar reasons,
Canizares et al (1993) concluded that for a given mass flux, the amount of X-ray emission from a 'heating flow' (where cool gas is
evaporated and implusively heated to the ambient cluster temperature)
is $\sim 10^{-4}$ times the emission from a cooling flow.}
  
In summary, the primary effect of a non-Maxwellian tail in the
electron velocity distribution in cluster gas is on the plasma
temperature rather than the ionization stage (except possibly for very low ionization stages, which have long
recombination times). Thus, the ionization stage of Fe remains a good
diagnostic of plasma temperature, and the lack of emission lines from
low ionization stages seems to suggest that the gas cannot cool below some cutoff temperature $T_{\rm crit}$.   

\section{Effects of Non-Thermal Pressure Support}

In this section, we very briefly consider the effects of a source of non-thermal
pressure support in cooling gas, either in the form of magnetic field
or cosmic rays. This will be relevant for evaluation of gas densities in our consideration of
photoionization in \S \ref{section:photoionization}. Observations of Faraday rotation in the cores of
clusters indicate the presence of highly chaotic $\ge 5-10 \mu$G
magnetic fields \cite{clarke2000}, which in some cases reach equipartition
values \cite{enblin1997}. In addition, the observation of synchrotron
emission, jet and radio galaxy activity mean that the energy density
in magnetic fields and cosmic rays could be considerable, and estimates show that it can be
comparable to the thermal energy density of the gas
\cite{enblin1997,volketal1996}. Finally, cosmic rays could also be
accelerated in gravitational shocks due to either accretion
\cite{miniati2001} or mergers \cite{donnellyetal2001}. As the gas cools,
the non-thermal component evolves adiabatically and can eventually
dominate the overall pressure support. Since the gas no longer cools
isobarically, this has two important consequences for us: (i) the
emission line spectrum differs from that of standard isobaric cooling
models; (ii) gas at low temperatures cools at lower densities, reducing the recombination rate
and allowing it to be more easily photoionized (see \S
\ref{section:photoionization}). In the limit where non-thermal
pressure dominates, cooling becomes isochoric rather than isobaric,
leading to a factor of at most $\sim (5/2)/(3/2) \sim 5/3$ reduction in
the strength of low temperature emission line features; thus effect
(i) by itself cannot account for the factor $\ge 3$ reduction in the
strength of low temperature emission lines. However, effect
(ii) implies that the recombination times could be longer by up to a factor $\sim
(T_{i}/T)$ from the isobaric cooling case (where $T_{i}$ is the initial background temperature, and
$T$ is the temperature at which a particular ionic stage becomes
abundant). This permits photoionization to dominate over recombination, and the latter effect 
could be quite important. Here we only consider the dynamical
effects of a non-thermal component; we shall consider the possible
implications of cosmic-ray heating or magnetic reconnection (when the
non-thermal component does not evolve adiabatically) in a subsequent paper. 

\begin{figure}
\psfig{file=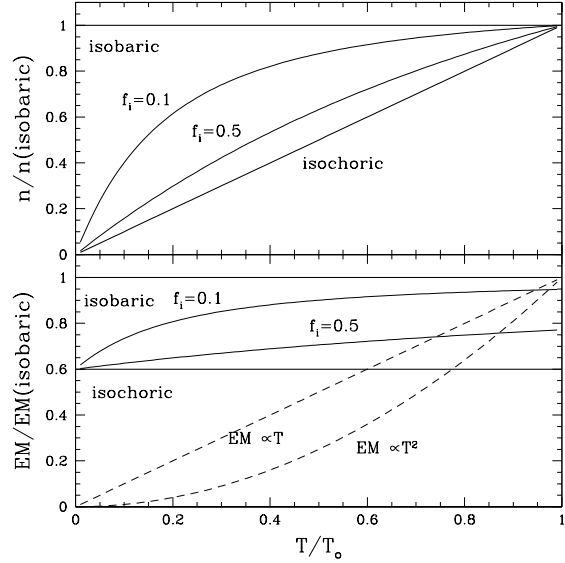,width=80mm}
\caption{{\it Bottom panel}: The emission measure as a function of
  temperature for models with initial fractional non-thermal pressure $f_{i}=
  P_{NT}/P_{tot}$ at temperature $T_{o}$. Nonthermal pressure support results
  in smooth interpolation between the isobaric and isochoric cooling
  regimes and cannot account for the observed emission measure, where
  $EM \propto T^{\alpha}$, with $\alpha \approx 1-2$. {\it Top panel}:
  Density as a function of temperature, normalized to the density the
  gas would have if it cooled isobarically. Non-thermal pressure
  become progressively more important at lower temperature, leading to significantly lower densities than if the
  gas cools isobarically. This could increase recombination times and
  delay the formation of low ionization stage ions, if there is a
  significant photoionizing flux (see \S \ref{section:photoionization}).}
\label{fig:nonthermal_plot}
\end{figure}

What is the effect of non-thermal pressure support on a thermally
unstable clump of gas? If we model the non-thermal pressure as a
relativistic gas with adiabatic index $\gamma$ (henceforth we shall
assume $\gamma=4/3$) then the total pressure:
\begin{equation}
n k_{B} T + K_{NT} n^{\gamma} = {\rm const}
\label{eqn:NT_pressure}
\end{equation}
is conserved during the cooling process, if the cooling time is much
shorter than the characteristic flow time, $t_{cool} \ll
t_{flow} \sim R/u$, which is generally the case (we have also assumed that the
magnetic field lines and any associated cosmic rays are frozen into the
plasma). Given the initial gas pressure, non-thermal pressure and
temperature of the gas, $P_{gas,i}, P_{NT,i},T_{i}$, we can solve
equation (\ref{eqn:NT_pressure}) for the density as a funtion of
temperature $n(T)$, assuming that the adiabatic constant
$K_{NT}=P_{CR,i}/n_{i}^{\gamma}$ is conserved. The gas density at given temperature is
always less than that assumed by the isobaric cooling model $n(T) <
n_{isobaric}=n_{i} T_{i}/T$; also, at low temperatures non-thermal pressure support
eventually dominates, and the gas cools almost isochorically. These
effects are illustrated in Figure \ref{fig:nonthermal_plot}, for different
values of the initial non-thermal pressure
$f_{i}=P_{NT,i}/P_{tot,i}$. For cluster
cores, it is reasonable to assume up to equipartition values ($f_{i}
\sim 0.5$) for the non-thermal pressure, before the onset of thermal
instability. We will use
equation (\ref{eqn:NT_pressure}) to compute the densities $n(T)$ when
calculating recombination rates in \S \ref{section:photoionization}. 

We can easily compute the effect of the non-thermal pressure support
on the emission measure of the gas as a function of temperature. It now
becomes:
\begin{equation}
\frac{d EM}{dT}(T_{2})= \left( \frac{3}{2} + \frac{n_{2}/n_{1}-1}{T_{1}/T_{2}-1} \right) \frac{\dot{M}}{\mu m_{p}}
\frac{k_{B}}{\Lambda(T)}. 
\end{equation}
where $n_{1},T_{1}$ are the initial temperature and density, and
$n_{2}(T_{2})$ is obtained from equation \ref{eqn:NT_pressure}. In the
limit where non-thermal pressure is unimportant,
$n_{2}/n_{1}=T_{1}/T_{2}$, and the prefactor $\rightarrow 5/2$, for
isobaric cooling; when non-thermal pressure completely dominates,
$n_{2} \approx n_{1}$ and the prefactor $\rightarrow 3/2$, for
isochoric cooling. 
%Comparison with equation
%\ref{eqn:emission_measure} shows that this does not adequately
%suppress the emission measure at low temperatures; however, it could
%%be important in tandem with other suppression mechanisms which work at the
%factor $\sim 2$ level.
 
Non-thermal pressure support will also retard gas inflow and reduce
the amount of work done on the cooling gas by the gravitational
potential. Note that the compression of gas in the dark matter potential will
make little difference to the overall gas cooling rate, since for
temperatures and densities characteristic of cooling flow clusters the
gas cooling time depends almost exclusively on its entropy. In particular, at fixed
entropy the cooling time is almost independent of the gas density;
thus, the total cooled gas mass is almost unchanged whether the gas is adiabatically
compressed by the gravitational potential or not \cite{ohbenson}. Thus, the
presence or absence of non-thermal pressure support makes little
difference in terms of the actual mass drop-out rate (although it can have a
large effect on the {\it inferred} mass drop-out rate, since the
bolometric X-ray luminosity of the cluster can be significantly reduced
when non-thermal pressure restricts inflow). However, its main effect
is in altering the normalization rather than the shape of the X-ray
spectrum \cite{petersonetal2002,nulsen98,allen2000}. In particular, compression by the gravitational potential
makes little difference to lines which appear at low temperatures,
since $t_{cool} \ll t_{flow} \approx R/c_{s}$, and the gas cools
before it travels a significant distance. As noted by
\scite{petersonetal2002}, even if there were no mass drop-out, the maximal
effect of gravitational inflow would be to replace in equation
\ref{eqn:emission_measure} (when $\alpha=0$) the factor $5/2
\rightarrow 3/2 -\lambda_{T}/\lambda_{\rho}$, where $\lambda_{T}$ is
the radial logarithmic temperature gradient and $\lambda_{\rho}$ is
the radial logarithmic density gradient; $\lambda_{T}/\lambda_{\rho}$
is always observed to be negative in the cores of clusters. 

\section{Photoionization}  
\label{section:photoionization}

\subsection{Is Photoionization of Metals in Clusters Plausible?}

From the preceding discussion in \S \ref{section:coulomb_heat} it is clear that if we wish to break the
correspondence between metal ionization stage and plasma temperature
by ionizing the metals, we need to use an ionizing source which
injects little heat per ionization. Photoionization is ideal for this
purpose: it only targets the metal ions (the Compton optical depth
$\tau \le 10^{-2}$ through clusters is small); and since metal ions
are a rare tracer species, photoheating is generally unimportant. Furthermore, in
clusters a large reservoir of hot X-ray emitting gas is available to
provide ionizing photons. In clusters it
is generally assumed that the plasma is in the coronal limit and that
the ambient radiation field is too weak to play a role in the
ionization properties of the gas. Below we perform a simple order of
magnitude estimate to show that in fact photoionization effects may be
non-negligible; indeed, it has been proposed \cite{donavoit91,voitetal94}
that the optical filaments seen in clusters are powered by
self-irradiation by the surrounding hot cluster gas. Once again, since
the free-free luminosity scales with the virial temperature of the
cluster, this would provide an explanation of the universal scaling of
the cut-off temperature with the cluster temperature: the harder
radiation field in hotter
clusters can induce photoionization of progressively higher ionization stages.    
For definiteness, we once again consider the ionization of
Fe XVII, an ion indicative of plasma cooling down to 0.2-0.8 keV,
which is notoriously absent in the spectra of cooling flow
clusters \cite{petersonetal2001}. In steady state,
the energy advected into a cooling region should balance the radiative
energy emerging. As \scite{voitetal94} argue, the former should be $\sim u(r) v(r)$, where $u(r)$ is the effective
energy density at $r$ and $v(r)$ is the inflow velocity. If the
cooling occurs isobarically, then $u(r) \approx 5P(r)/2$. The inflow
speed should be of order the sound speed in the gas, $v(r) \approx
c_{s} \sim 300 {\rm km s^{-1}}$, where the latter is the isothermal sound
speed of gas at $\sim 1$keV. This energy flux $\sim u(r) v(r)$ is a reasonable estimate of the ambient radiation field in a cluster. If it mostly emerges in the form of free-free emission, the rate at which a metal atom is
photoionized is:
\begin{equation}
R_{ion} \approx \frac{5}{2} \frac{P c_{s}}{E_{i}} f_{i} \sigma_{i} 
\end{equation}
where $E_{i}$ is the energy of the ionization edge, $\sigma_{i}$ is
the photoionization cross-section at $\sim E_{i}$, and $f_{i}$ is a
correction factor for the fraction of the flux emerging above $E_{i}$;
for a free-free spectrum typically $f_{i}$ is of order unity. By contrast, the recombination rate is:
\begin{equation}
R_{\rm rec} = \alpha(T) n_{e} = \alpha(T) \left( \frac{P}{k_{B}T} \right)
\end{equation}
where $\alpha(T)$ is the temperature dependent recombination
coefficient, and we have assumed that cooling occurs isobarically. The
ratio of the photoionization rate to recombination rate is therefore:
\begin{equation}
{\rm \frac{Photoionization}{Recombination}} \approx  \frac{k_{B}
  T}{E_{i}}  \left( \frac{c_{s} \sigma_{i}}{\alpha(T)} \right)
  \sim 0.3  \left( \frac{k_{B} T}{E_{i}} \right) \left(
  \frac{\alpha(k_{B} T)}{\alpha(E_{i})} \right) 
\end{equation}
where the quantities $\alpha \sim 10^{-11} {\rm cm^{3} s^{-1}}$,
$\sigma_{i} \sim 10^{-19} {\rm cm^{-2}}$ have been evaluated for $k_{B}
T \sim E_{i} \sim 1.26 {\rm keV}$, as is appropriate for Fe XVII. It is
interesting to note that this result is independent of the pressure
and hence the density of the gas: for isobaric conditions, higher
densities result in a higher recombination rate but also a higher
photoionization rate, since the surrounding gas has a higher
emissivity. The close match of the photoionization and recombination
rates in this rough estimate warrants further investigation. In addition,
we shall see that other sources of photoionizing radiation such as AGN
or X-ray binaries could be even more important.

\subsection{Estimating the Ambient Radiation Field}

Let us quantify the level of photoionization needed to explain the
observational results. We define a critical flux $F_{crit}$, such that the
photoionization rate $R_{ion}=\int dE F_{crit}(E) \sigma_{i}(E)$ is roughly
equal to the recombination rate to Fe XVII, $R_{rec}(T_{c})=\alpha(T_{c}) n_{e,c}$ where $n_{c}$ is the electron number
density at $T_{c}=0.4 {\rm keV}$, the temperature at
which the abundance of the Fe XVII ion peaks. For free-free emission
from a plasma at temperature $T_{h} \approx 5$keV, the critical flux
at the ionization edge $E_{i}=1.26 {\rm keV}$ of Fe XVII is:
\begin{equation}
F_{crit}(E_{i})= 4.4 \times 10^{5} \left( \frac{n_{c}}{0.1 {\rm 
    cm^{-3}}} \right) {\rm keV cm^{-2} s^{-1} keV^{-1}}
\end{equation}
Due to the flat spectrum of free-free emission up to $\sim T_{h}$,
this depends only very weakly on the temperature of the hot plasma
$T_{h}$, as long as $T_{h} \gg E_{i}=1.4$keV. For a power law ionizing
spectrum $F \propto E^{-\Gamma}$ where $\Gamma=0.7$, we
obtain a very similar result, $F_{crit}^{AGN}(E_{i})= 4.2 \times
10^{5} ( {n_{c}}/{0.1 {\rm cm^{-3}}} ) {\rm keV
  cm^{-2} s^{-1} keV^{-1}}$. This gives us a rough estimate of when
photoionization is non-negligible; henceforth, we will normalize all our
results to $F_{crit}(E_{i})$. 

Let us first consider the effects of free-free emission from hot
cluster gas. We estimate the internal radiation field of a cluster following \scite{sazonovetal}, who did so in order to constrain resonant X-ray scattering. Consider a point some distance $r$ from the center of a
cluster. Integrating over all solid angle, the total flux of radiation received at that point is:\begin{equation}
F(r)=\int_{-1}^{1} d\mu  \int_{0}^{\infty} dr^{\prime} \epsilon(R)
\label{eqn:flux}
\end{equation}
where $\epsilon(r)$ is the emissivity profile of the gas, and
$R^{2}=r^{2}+r^{\prime 2} - 2 r r^{\prime}\mu$. The continuum emissivity of cluster gas in bremsstrahlung is:
\begin{eqnarray}
\epsilon(E)=2.3 \times 10^{-20} T^{-1/2} {\rm exp}(-E/k_{B}T)
n_{e}^{2} g_{B}(T,E) \\ \nonumber {\rm erg \, cm^{-3} s^{-1}
  keV^{-1}}.
\label{eqn:free_free}
\end{eqnarray}
where $g_{B}(T,E) \approx (E/k_{B} T)^{-0.4}$ is the Gaunt factor. Let us consider a cluster with a beta-law radial density profile:
\begin{equation}
n_{e}=\frac{n_{o}}{(1+r^{2}/r_{c}^{2})^{3\beta/2}}
\label{eqn:beta_profile}
\end{equation}
where $n_{o}$ is the central electron density and $r_{c}$ is the core
radius. Equation (\ref{eqn:flux}) can be integrated numerically to
obtain the X-ray flux at any point in the cluster. However, there are
two interesting analytic limits. At the center of the cluster: 
\begin{eqnarray}
&&F(r=0,E)=10^{7} \frac{\Gamma(3\beta-1/2)}{\Gamma(3\beta)}
\centralhi^{2} \left( \frac{r_{c}}{100 {\rm kpc}}\right)
\\ \nonumber &&\times \left( \frac{k_{B} T}{5 \, {\rm keV}}\right)^{-1/2} x^{-0.4}
  {\rm exp}(-x) \ \ {\rm keV \, cm^{-3} s^{-1}
  keV^{-1}} \\ \nonumber
&&\approx 3.2 F_{crit} (E_{i}) \left( \frac{r_{c}}{100 {\rm kpc}}\right)
\centralhi \left( \frac{n_{c}/n_{o}}{5}\right)^{-1}
\\ \nonumber \times &&\left(\frac{E}{E_{i}} \right)^{-0.4} {\rm
  exp}[-x+x_{i}] 
\end{eqnarray}
where $x=E/k_{B}T$. Also, for the special case of $\beta=2/3$, the flux traces the density profile exactly:
\begin{equation}
F(r,E)=F(0,E) \frac{1}{1+(r/r_{c})^{2}} \propto n(r)
\end{equation}
Since $F_{crit} \propto n$, this implies that $F/F_{\rm crit}=$
const throughout the cluster, i.e., the decreasing flux and
photoionization rate at the outer parts of the cluster gas is tracked
exactly by the decreasing densities and hence decreasing recombination
rate. For gas in an isothermal potential, this implies that photoionization
suppression of lower ionization stages is constant throughout the
cluster. Note that, even if the assumption of isothermality is
dropped, $F \propto n$ is a very good approximation (since the
emissivity depends largely on density and is only weakly temperature
dependent). This could provide a natural explanation as to why the
cluster gas appears locally isothermal at each radius \cite{molpiz,matsu}: even if
the cluster gas is multi-phase, the cooler gas is overionized for its
electron temperature, and thus the emission measure of lower
ionization stage ions indicative of cool gas is unmeasurably small. Furthermore, this effect is relatively independent of cluster radius.

Other possible sources of X-ray ionizing photons are AGN and X-ray
binaries. The flux from an AGN is $F_{\rm AGN}=\frac{(2-\alpha)L_{X}}{4 \pi
  (E_{2}^{2-\alpha}-E_{1}^{2-\alpha}) r^{2}} E^{1-\alpha}$, or numerically:
\begin{eqnarray}
F_{\rm AGN}&=& 1.2 \times 10^{8} \left( \frac{M_{BH}}{10^{9} M_{\odot}} \right)
  \left( \frac{L_{X}/L_{\rm Edd}}{0.05} \right) \left( \frac{r}{10
  {\rm kpc}} \right)^{-2} \\ \nonumber  && \left( \frac{E}{1 {\rm keV}} \right)^{-0.7}
{\rm keV \, cm^{-3} s^{-1} keV^{-1}}   
\end{eqnarray} 
where $\alpha \approx 1.7$ is the photon index, $L_{X}$ is the AGN
luminosity in the energy range $E_{1}-E_{2}\approx 1-10$keV range, and we
have assumed that $\sim 5\%$ of the bolometric luminosity emerges in
this range \cite{elvis1994}. Assuming again a beta-law profile for the
density, we find that:
\begin{eqnarray}
&&F_{AGN}(E)  \approx 50 F_{\rm crit}(E_{i}) \left( \frac{M_{BH}}{10^{9}
  M_{\odot}} \right) \left( \frac{L_{X}/L_{\rm Edd}}{0.05} \right)
  \centralhi^{-1} \\ \nonumber && \left[ \left( \frac{r_{c}}{100 {\rm
  kpc}} \right)^{-2} + \left( \frac{r}{10
  {\rm kpc}} \right)^{-2} \right] \left( \frac{n_{c}/n}{5} \right)
  \left( \frac{E}{{\rm 1 keV}} \right)^{-0.7}
\end{eqnarray}

It is therefore apparent that particularly near the center (where soft
X-ray emission lines are most egregiously missing), an X-ray
bright AGN could produce photo-ionizing radiation orders of magnitude
higher than required to dominate collisional processes. It is also
interesting to note that since $F_{AGN} \propto r^{-2}$ and $F_{\rm crit} \propto n
\propto r^{-2}$, outside the central density core the AGN flux is
also always some constant multiple of the critical flux, similar to
the free-free radiation case. The invocation
of AGN to provide ionizing radiation rather than the ambient hot
cluster gas is somewhat less desirable, because such sources have a short duty
cycle and are not seen in all cooling flow clusters (although Thompson and resonant line
scattering could trap their radiation for some time and boost the
X-ray surface brightness by a factor $\sim 3-10$ over the free-free
continuum \cite{sazonovetal}. Furthermore, the universal scaling of
the cut-off temperature with the ambient plasma temperature is
lost, although one could argue that $F_{AGN} \propto M_{bh} \propto
\sigma_{CD}^{\alpha} \propto T_{cluster}^{\alpha/2}$, where $\sigma_{CD}$ is the
velocity dispersion of the CD galaxy and $\alpha$ is the slope of the
$M_{bh}-\sigma$ relation. Nonetheless, it is
clear that photoionization effects could be important. In the
following section we consider if a photo-ionizing flux of arbitrary
magnitude could be responsible for the observed emission line
spectrum. 

\subsection{Ionic Abundances and Line Suppression}
\label{section:cloudy}

Let us now perform a more careful calculation of the effects of
photoionization on the emission spectrum of cooling gas. Since $t_{\rm
  rec} \ll t_{\rm cool}$, the assumption that the metal ionization
stages assume their equilibrium values (where photoionization and
collisional ionization balance recombinations) is a good one. However,
when computing the emission line spectra, several effects need to be
taken into account. One is that the photoionization cross section
take into account Auger multi-electron ejection. This process
increases the impact of photoionization of inner shell electrons (as
many as 8 electrons can be removed from the photoionization of the 1s
shell), and couples non-adjacent stages of ionization, which makes it
necessary to iterate on the ionization solution. The second is
the change in the cooling function due to the deviation of the ionic
abundances from their collisional equilibrium values.  

\begin{figure}
\psfig{file=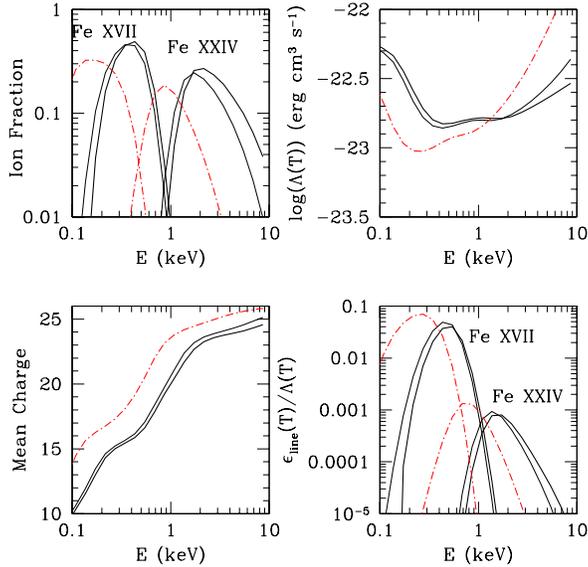,width=80mm}
\caption{The ionic fraction, mean charge, cooling function, and
  cooling-function weighted emissivity for the FeXVII $15 {\rm \AA}$ line
  and FeXXIV $11 {\rm \AA}$ line for isobarically cooling gas
  subject to a photo-ionizing radiation field $F_{AGN} \propto
  \nu^{-0.7}$ of strength 0 (dark solid line), $F_{\rm crit}$
  (indistinguishable from no radiation), $10 F_{\rm crit}$ (light
  solid line) and $100 F_{\rm crit}$ (dot-dashed line). Only at very
  high intensities does the photoionizing background have significant
  impact. The overall effect of an ionizing background is to increase
  the mean charge of a system at all temperatures. Although there are more soft
  photons able to photoionize lower ionization stages, at lower
  temperatures, densities and hence recombination rates are much
  higher, and the two effects roughly cancel. The results of
  calculations with a free-free ionizing background with equivalent values of $F_{\rm crit}$ do not
  differ significantly.}
\label{fig:vary_Fcrit}
\end{figure}

\begin{figure}
\psfig{file=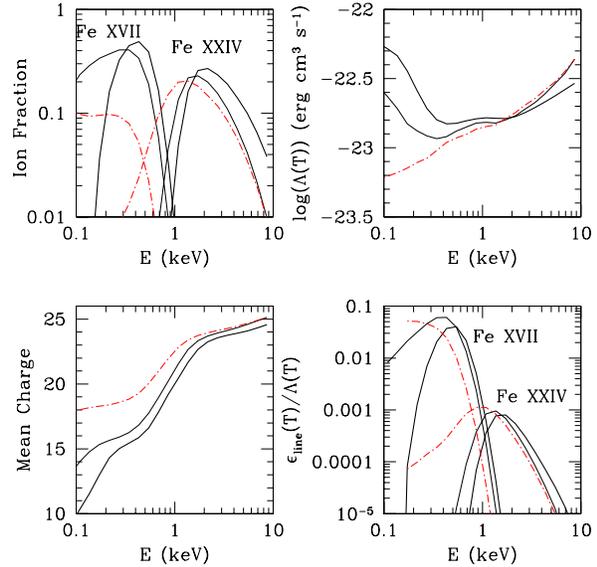,width=80mm}
\caption{The ionic fraction, mean charge, cooling function, and
  cooling-function weighted emissivity for the FeXVII $15 {\rm \AA}$ line
  and FeXXIV $11 {\rm \AA}$ line for gas where only collisional process
  operate and is isobarically cooling (dark
  solid line) and gas subject to a AGN radiation field
  of strength $10 F_{\rm crit}$ which has initial non-thermal pressure
  support $P_{NT} \sim 0.1 P_{tot}$ (light solid line) and $P_{NT} \sim 0.5 P_{tot}$
  (dot-dashed line). Because a non-thermal component causes the gas to
  cool isochorically at low temperature, densities and recombination
  rates are significantly lower at lower temperatures, and X-ray
  photoionization can become significantly more important than
  before. Naively, we would expect from the suppression of FeXVII
  abundance down to lower temperatures (upper left
  panel) that the emission measure of Fe XVII is reduced. However,
  because of the accompanying reduction in the cooling function at low
  temperature (upper right), the emission measure of Fe XVII integrated over all temperatures, is in
  fact {\it increased} (lower right).}
\label{fig:vary_NT}
\end{figure}

We therefore run CLOUDY \cite{cloudy}, which can compute both of
these effects. We present the results of these calculations in Figs
\ref{fig:vary_Fcrit} and \ref{fig:vary_NT}. In Figure
\ref{fig:vary_Fcrit}, we show the result of photoionization on
isobarically cooling gas, for different values of the radiation
field. Only large values $\sim 100 F_{\rm crit}$ of the radiation
field have a significant impact. Furthermore, the net effect of the
X-ray background is simply to increase the mean charge of the system
at a given temperature, and shift the appearance of lower ionization
stages to lower temperatures. Naively, one might have expected the
X-ray background to become increasingly important at low temperatures, since lower
ionization stages have lower ionization thresholds (and thus are more
easily ionized). However, the greater densities and recombination
rates at lower temperatures in isobarically cooling gas roughly offset
the increase in the photoionization rate. Note that a photoionizing radiation field {\it decreases} the amplitude of the
cooling function at low temperatures. This can be easily understood:
because the metal ions are over-ionized with respect to their purely
collisional values, the relatively cold electrons cannot excite
atomic transitions which result in cooling radiation. In the limit
where the photoionization is very large and this effect dominates,
then line cooling is no longer important and cooling is entirely due
to bremmstrahhlung, $\Lambda(T) \propto T^{1/2}$. We have checked directly that photo-heating is entirely
negligible in this temperature range, even for large ($\sim 100 F_{\rm
  crit}$) radiation fields. The most important quantity is
$\epsilon_{\rm line}(T)/\Lambda(T)$ (lower right panel), the cooling
function weighted emissivity of a given line; the integral of this
function over all temperatures is directly proportional to the
emission measure of the line, equation (\ref{eqn:line_luminosity}). It
is this quantity we wish to suppress in lines associated with lower
ionization stages such as Fe XVII. Overall, we see that for isobarically cooling gas,
this quantity is not significantly suppressed, even for high levels of
the radiation field. Results for a radiation field normalized to the
same value of $F_{\rm crit}$ but for a free-free spectrum yield very
similar results. 

In Figure (\ref{fig:vary_NT}), we show the effects of non-thermal
pressure support, for a AGN ionizing background of $10 F_{\rm crit}$,
and different levels of initial nonthermal pressure support
$f_{i}=P_{NT}/P_{tot}$. Since the effects of photoionization scale as
$F_{\nu}/n$, if non-thermal pressure support impedes compression of
the gas to high densities, then (particularly at low temperatures) the
effects of photoionization could become considerably more
important. We see that this is indeed the case; the appearance of
lower ionization stages is suppressed at low temperatures. If there
were no change in the cooling function, then the emission measure of
lines associated with low ionization stages such as FeXVII would be
reduced by a factor of a few (depending on the strength of the
ionizing radiation field), as required to explain the observations. However, the reduction of the
cooling function at low temperatures noted above offsets this effect
. The net result is that the emission measure of lines associated with
low ionization stages is comparable or even {\it larger} than the case
where there is no ionizing radiation. Overall, it is not possible to decouple the cooling and ionization
properties of the gas because the emission lines we wish
to suppress appear at temperatures when metal line cooling dominates. Thus, this mechanism
would only be important in gas where the cooling function is
independent of the ionization state of the metals; for instance, in
very low metallicity gas where free-free emission cooling dominates.

\section{Conclusions}

In this paper we have examined two means of decoupling the ionization
stage of metal ions from the plasma temperature: excess ionization
from a non-Maxwellian tail of suprathermal electrons, and
photo-ionization either from free-free emission by hot cluster gas or
from internal AGN or X-ray binaries. These are very attractive because
they should be universal and also become important at some fixed
fraction of the ambient plasma temperature, observed properties which
none of the present-day models based on gas heating can
explain. Neither of these are successful: suprathermal electrons are much more efficient at heating the gas than
ionizing the metals, and photo-ionizing radiation decreases gas
cooling at low temperatures, ultimately {\it increasing} the emission
measure of emission lines associated with low temperature
gas. Nonetheless, they have attractive features which may prove useful
in elucidating a final solution. In a
companion paper, we shall examine the feasibility of solutions involving heating. In particular, we shall examine the global
stability of cooling flow clusters, addressing the classic cooling
flow problem of mass dropout. 

\section*{Acknowledgements}

I thank Roger Blandford and John Peterson for helpful conversations.

\end{document}